# Full-Blind Delegating Private Quantum Computation

Wen-Jie Liu[1,2,*], Zhen-Yu Chen[2], Jin-Suo Liu[3], Zhao-Feng Su[4], and Lian-Hua Chi[5]

**Abstract:** The delegating private quantum computation (DQC) protocol with the universal quantum gate set $\{X, Z, H, P, R, CNOT\}$ was firstly proposed by Broadbent et al., and then Tan et al. tried to put forward an half-blind DQC protocol (HDQC) with another universal set $\{H, P, CNOT, T\}$. However, the decryption circuit of *Toffoli* gate (i.e., *T*) is a little redundant, and Tan et al.'s protocol exists the information leak. In addition, both of these two protocols just focus on the blindness of data (i.e., the client's input and output), but do not consider the blindness of computation (i.e., the delegated quantum operation). For solving these problems, we propose a full-blind DQC protocol (FDQC) with quantum gate set $\{H, P, CNOT, T\}$, where the desirable delegated quantum operation, one of $\{H, P, CNOT, T\}$, is replaced by a fixed sequence $(H, P, CZ, CNOT, T)$ to make the computation blind, and the decryption circuit of *Toffoli* gate is also optimized. Analysis shows that our protocol can not only correctly perform any delegated quantum computation, but also holds the characteristics of data blindness and computation blindness.

## 1. Introduction

Blind quantum computation (BQC) is a novel model of quantum computation, where the client with limited quantum resources can perform quantum computation by delegating the computation to an untrusted quantum server, and the privacy of the client can still be guaranteed. As BQC provide a convenient and safe way to access the quantum computation, it may be an ideal model for the quantum application in the early days of "quantum computer era".

BQC can be generally divided into two categories: one is the measurement-based blind quantum computation (MBQC), and the other is the circuit-based blind quantum computation (CBQC). In MBQC, measurement is the main driving force of computation, which follows the principle of "entangle-measure-correct", and a certain number of quantum qubits are entangled to form a standard graph state. To be specific, it first prepares a certain graph state according to the requirements of desirable computation, and measures the first qubit according to the computation, then the measurement result will decide the following measurement basis which is known as "correction". In 2009, Broadbent et al. [Broadbent, Fitzsimons and Kashefi (2008)] proposed the first MBQC protocol, where the client generates the rotated single photons, and he sends them to the server to build the brickwork state that can implement the specific quantum computation. Since then, some other MBQC protocols were proposed [Morimae (2012a); Morimae (2012b); Li (2014); Xu (2014); Morimae (2015); Kong (2016)].

[1] Jiangsu Engineering Center of Network Monitoring, Nanjing University of Information Science & Technology, Nanjing 210044, China.

[2] School of Computer and Software, Nanjing University of Information Science and Technology, Nanjing 210044, China.

[3] State Grid Electric Power Research Institute, NARI Group Corporation, Nanjing 210003, China.

[4] Centre for Quantum Software and Information, Faculty of Engineering and Information Technology, University of Technology Sydney, NSW 2007, Australia.

[5] Department of Computer Science and Information Technology, La Trobe University, VIC 3086, Australia.

[*] Corresponding Author: Wen-Jie Liu. Email: wenjiel@163.com.

Different from MBQC, CBQC is based on the traditional circuit, which can be composed of all kinds of quantum gates. In 2005, Childs [Childs(2005)] proposed the first CBQC protocol based on the ideal of encrypting data with quantum one-time pad [Ambainis(2000); Boykin (2003)], however the client must possess quantum memory and the ability to execute the quantum SWAP operation. And in 2006, Arrighi and Salvail [Arrighi and Salvail (2006)] proposed another CBQC protocol for the calculation of certain functions, i.e., not the universal quantum computation, and it requires Alice to prepare and measure multi-qubit entangled states. Since then, some other CBQC protocols [Aharonov(2008); Broadbent (2013)] have been proposed. Recently, the concept of delegating private quantum computation has been proposed, which belongs to the CBQC model. In 2015, Broadbent [Broadbent (2015)] proposed the first delegating private quantum computation (DQC) protocol with the universal quantum set $\{X, Z, H, P, R, CNOT\}$. In one way, they relax the requirements of fully homomorphic encryption [Rivest (1978); Gentry (2009)] by allowing interaction. But at the same time, they strengthen the requirements by asking for information-theoretic security. Later, Tan et al. [Tan and Zhou (2017)] proposed a half-blind DQC protocol (HDQC) with another universal set $\{H, P, CNOT, T\}$, where "half-blind" means that the server cannot learn anything about client's input and output (also referred as the blindness of data), but client's computation are exposed to the server, i.e., the blindness of computation cannot be guaranteed. Obviously, the half-blindness of quantum computation is undesirable, because the privacy of computation is also an important aspect of information security.

Compared with previous works, the main contribution of our work is to propose a full-blind DQC protocol (FDQC) with universal gate set $\{H, P, CNOT, T\}$, where the desirable delegated quantum operation, one of $\{H, P, CNOT, T\}$, is replaced by a fixed sequence $(H, P, CZ, CNOT, T)$ to make the computation blind. In addition, we also optimize the decryption circuit of *Toffoli* gate, and also solve the problem of information leak in Tan et al.'s protocol. The rest of this paper is organized as follows. In the following section, we briefly review DQC and HDQC. A full-blind delegating private quantum computation protocol with universal gate set $\{H, P, CNOT, T\}$ is proposed in Sect. 3, and the correctness and full-blindness are discussed in Sect. 4. Finally, conclusion is drawn in the last section.

**2. Review of DQC and HDQC**

*2.1. Review of DQC*

DQC enables an almost-classical client to delegate the execution of any quantum computation to a remote server without exposing his information. The brief process of DQC [Broadbent(2015)] is as follows (also shown in Fig. 1).

- **Quantum encryption:** Client uses Pauli operations $X$ and $Z$ to encrypt quantum state $|\varphi\rangle$, and then obtains $|\varphi\rangle_{enc} = X^a Z^b |\varphi\rangle$, where $a$ and $b$ are the encryption keys randomly selected from $\{0,1\}$, after that he sends $|\varphi\rangle_{enc}$ to the server QC.
- **Quantum computation:** QC implements the specific quantum computation (certain unitary operation $U$) on the encrypted quantum state $|\varphi\rangle_{enc}$.
- **Quantum decryption:** The server returns the output state $U|\varphi\rangle_{enc}$ to the client. Then the client decrypts the output state: $X^{a'} Z^{b'} \left( U X^a Z^b |\varphi\rangle \right) \to U|\varphi\rangle$ according to the decryption rules, and finally gets the quantum computation result $U|\varphi\rangle$.

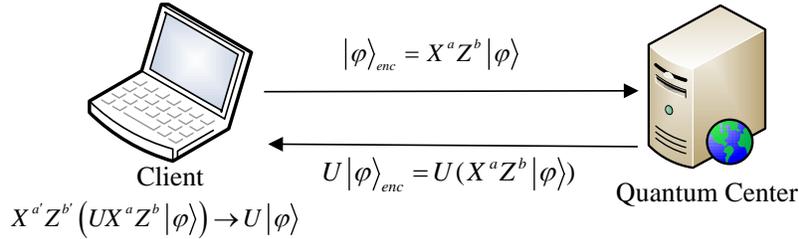

Client $\quad X^{a'} Z^{b'} \left( U X^a Z^b |\varphi\rangle \right) \to U|\varphi\rangle \quad$ Quantum Center

**Figure 1:** The main process of DQC model

As we know, the quantum gate set $\{X, Z, H, P, R, CNOT\}$ is universal [Nielsen and Chuang (2011)], which means it can be used to construct arbitrary quantum computation (i.e., arbitrary unitary operation $U$). These quantum gates have following properties,

$$\begin{cases} X|j\rangle = |j+1\rangle \\ Z|j\rangle = (-1)^j |j\rangle \\ H|j\rangle = \left(|0\rangle + (-1)^j |1\rangle\right)/\sqrt{2} \\ P|j\rangle = (i)^j |j\rangle \\ R|j\rangle = \left(e^{i\pi/4}\right)^j |j\rangle \\ CNOT|j\rangle|k\rangle = |j\rangle|j \oplus k\rangle \end{cases} \quad j \in \{0,1\} \tag{1}$$

And their encryption and decryption circuits are shown in Fig. 2.

## 2.1. Review of HDQC

Besides $\{X, Z, H, P, R, CNOT\}$, $\{H, P, CNOT, T\}$ is another discrete universal quantum gate set[Nielsen and Chuang (2011)], where $T$ is the *Toffoli* gate. As we know, the *Toffoli* gate is a reversible quantum gate which is more frequently used for constructing large-scale and complex quantum circuits. Recently, Tan et al. [Tan and Zhou (2017)] proposed a half-blind DQC protocol with $\{H, P, CNOT, T\}$, and its main contribution is to give the encryption and decryption circuit of the $T$ gate. As the encryption and decryption circuits of $H$, $P$ and $CNOT$ gates are the same as Broadbent's protocol, here we just review the encryption and decryption of *Toffoli* gate. During the encryption process, the client encrypts the first qubit with unitary operation $X^a Z^b$, the second qubit with $X^c Z^d$ and the third qubit with $X^e Z^f$, where the encryption keys

$a, b, c, d, e, f \in \{0, 1\}$ are randomly generated by the client. After the *Toffoli* gate performed by

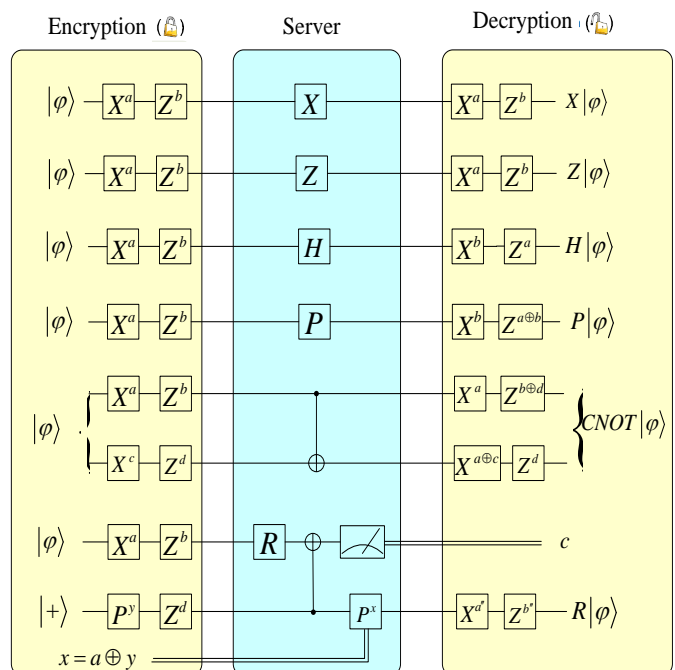

**Figure 2:** The encryption and decryption circuit for the universal quantum gate set $\{X, Z, H, P, R, CNOT\}$ [Fisher (2013)]

the server, the client cooperates with the server to perform the decryption with the extra *CZ*, *CNOT* and *SWAP* unitary operations as correction. The whole encryption and decryption circuit of *Toffoli* gate is shown in Fig. 3.

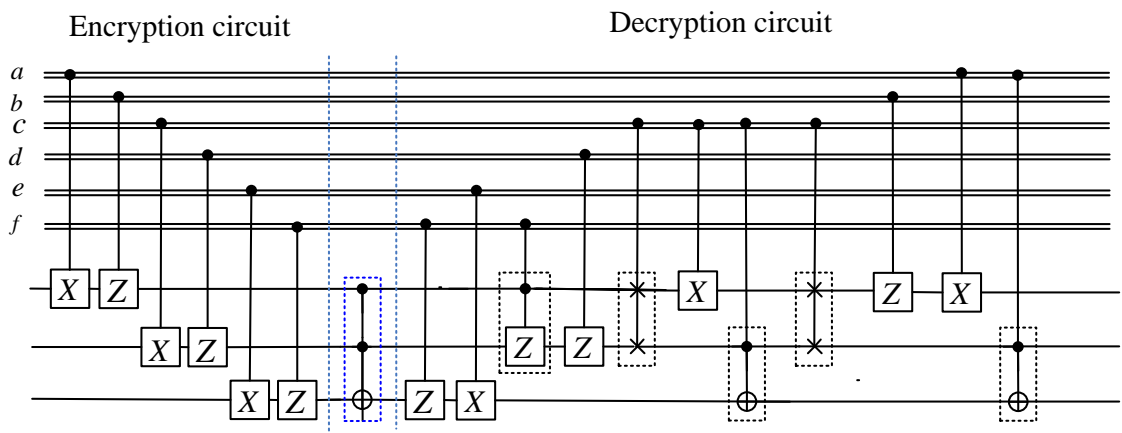

**Figure 3**: The encryption and decryption circuit of *Toffoli* gate. The quantum operations in the dotted box denote that they are performed in the server side.

However, there are two flaws in Tan et al.'s HDQC protocol. First, the protocol is half-blind, i.e.,

it only guarantees the blindness of the data. Although the server cannot get any information about the data, the desirable computation can be obtained by the server because the delegated operations are exposed to the server. Second, the protocol exists the information leak. To be specific, if the server is performing the *Toffoli* gate, the client may delegate the server to perform some correction operations. Referring to the decryption circuit in Fig. 3, the corrections are related with the encryption keys, i.e., the *CZ*, *SWAP* and *CNOT* corrections represent the encryption keys *f=1*, *c=1* and (*a=1, c=1*), respectively. Since the server knows all the delegated quantum operations, he can deduce the corresponding encryption key based on the above rules. For example, in the HDQC process for *Toffoli* gate, if the client asks the server to perform a *CZ* operation between the first and second qubit, then the secret key *f=1* will be revealed to the server.

## 3. Full-blind delegating private quantum computation

### 3.1. The FDQC protocol

Suppose that the client delegates the server to implement a certain quantum computation which is composed of quantum operations (i.e. quantum gates in $\{H, P, CNOT, T\}$), the procedures of FDQC are given as follows.

1. The client generates a 9-qubit sequence S which consists of ancillary qubits and message qubits. And then he divides the sequence into five subsequences (The first subsequence $S_H$ consists of the first qubit, the second subsequence $S_P$ consists of the second qubit, the third subsequence $S_{CZ}$ consists of the third and fourth qubits, the fourth subsequence $S_{CNOT}$ consists of the fifth and sixth qubits, the fifth subsequence $S_T$ consists of the remaining three qubits). It should be noted that the five subsequences $S_H$, $S_P$, $S_{CZ}$, $S_{CNOT}$ and $S_T$ are prepared for the fixed ordered operations $(H, P, CZ, CNOT, T)$.

2. According to the delegated quantum operation, the client chooses one of $\{S_H, S_P, S_{CZ}, S_{CNOT}, S_T\}$ as the message part (also called the message subsequence), and the other subsequences as the ancillary part (which will be used to confuse the delegated operation). For example, if the delegated quantum operation is *T* gate, then he chooses $S_T$ as the message part, and the remainder $(S_H, S_P, S_{CZ}, S_{CNOT})$ is the ancillary part.

3. The client encrypts every qubit $|\varphi\rangle$ in message subsequence by the unitary operation $X^a Z^b$, where $(a, b)$ are the encryption keys randomly chosen by him, $a, b \in \{0, 1\}$. After that, he sends the sequence *S* to the server.

4. The server performs the operations $(H, P, CZ, CNOT, T)$ on the qubits in subsequence $\{S_H, S_P, S_{CZ}, S_{CNOT}, S_T\}$, and returns the output qubits to the client.

5. The client extracts the message qubits from the output qubits according to his original selection in Step 1, and decrypts them by $X^{a'}Z^{b'}$ with the decryption keys $(a',b')$ (the decryption keys and decryption process will be detailed discussed in the next subsection). If the delegated quantum operation is T and the encryption keys $f=1$, $c=1$ or $a=1$, then the client will perform the corrections according to the following rule: if $f=1$, then correction operation is *CZ* between the first and second qubits; if $a=1$, then correction operation is *CNOT* between the first and third qubits; if $c=1$, then correction operation is *CNOT* between the second and third qubits. Then the client delegates the server to perform the correction operation as the above steps.

6. The client and the server repeat the above steps until all the delegated quantum operations are completed.

For ease of understanding, we suppose the client wants to delegate the operation

$$U = \begin{bmatrix} \frac{1}{\sqrt{2}} & \frac{1}{\sqrt{2}} \\ \frac{i}{\sqrt{2}} & -\frac{i}{\sqrt{2}} \end{bmatrix} = \frac{1}{\sqrt{2}}\begin{bmatrix} 1 & 0 \\ 0 & i \end{bmatrix}\begin{bmatrix} 1 & 1 \\ 1 & -1 \end{bmatrix} = PH \qquad (2)$$

to the server, and Fig. 4. describes the whole delegation process of *U*.

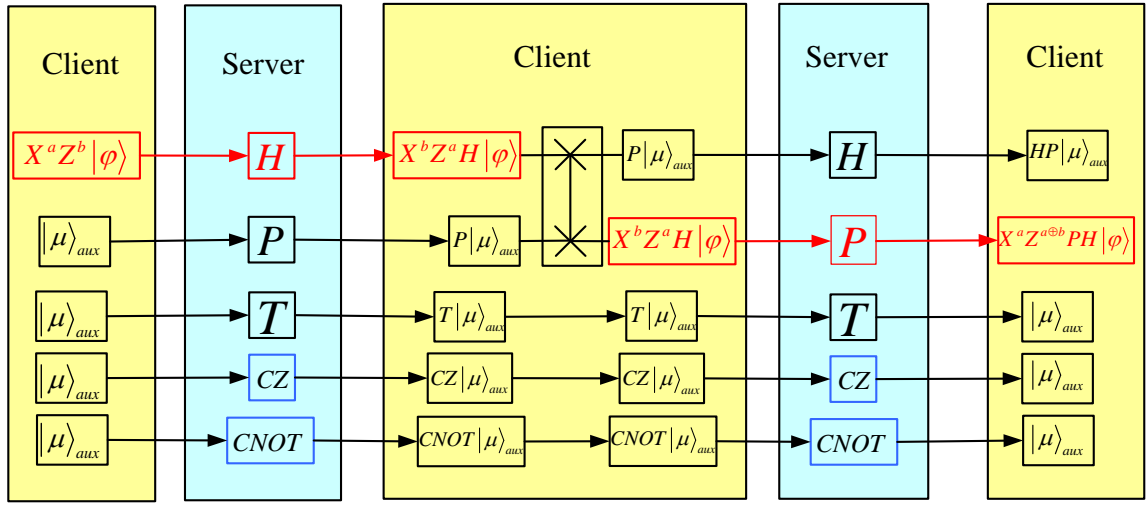

**Figure 4**: The delegation process of *U = PH* in our protocol. $|\mu\rangle_{aux}$ is ancillary qubit randomly generated by the client. The operations in dotted box are the desirable operations.

In the proposed FDQC protocol, the fixed order operations $(H, P, CZ, CNOT, T)$ are all performed in each round, which will confuse the delegated quantum operation and finally achieve the computation blindness.

### *3.2. The encryption and decryption of universal quantum gate set*

As we know, both $\{X,Z,H,P,R,CNOT\}$ and $\{H,P,CNOT,T\}$ are the universal quantum gate sets that can construct arbitrary quantum computation. Compared with other quantum gates, the T gate (i.e., *Toffoli* gate) is used more often as a basic unit for constructing large complex circuits. So, we choose the universal quantum set $\{H,P,CNOT,T\}$ to implement the FDQC protocol.

Through in-depth analysis of the relevant characteristics of *H*, *P*, *CNOT* and *Toffoli* gates (partly shown in Equation (1)), we derive the relevant quantum homomorphic decryption method for these gates, and further give the encryption and decryption process for the quantum set $\{H,P,CNOT,T\}$ in the FDQC protocol. All the encryption and decryption processes for these gates can be sketched in Figs. 5-8.

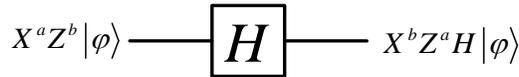

**Figure 5**: The encryption and decryption process for *H*.

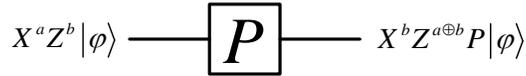

**Figure 6**: The encryption and decryption process for *P*.

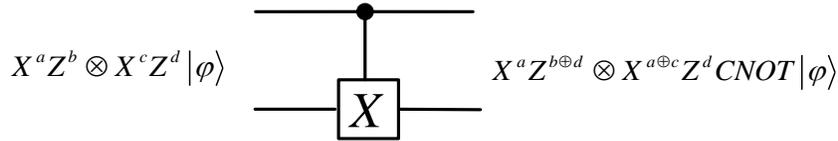

**Figure 7**: The encryption and decryption process for *CNOT*.

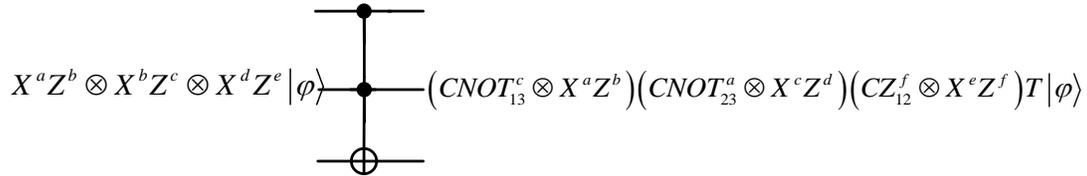

**Figure 8**: The encryption and decryption process for *T* (i.e., Toffoli gate).

Since the encryption and decryption circuits of $H,P,CNOT$ are given in [Broadbent (2015)] and [Tan and Zhou (2017)], here we skip them and focus on the description of *Toffoli* gate. In our study, we simplify the encryption and decryption circuit of *Toffoli* (shown in Fig. 9).

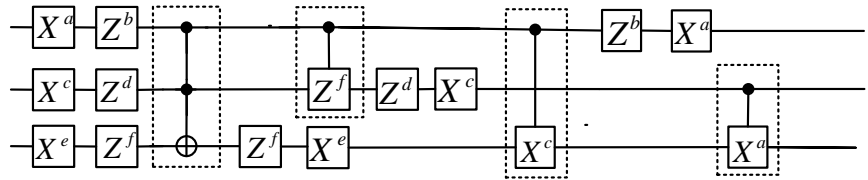

**Figure 9**: The encryption and decryption circuit of *Toffoli gate*. Here, the two-qubit operations in dotted box are viewed as correction.

Assume that the encryption keys $(a,b),(c,d),(e,f)$ for the encryption circuit of *Toffoli* gate is randomly generated by the client, he encrypts the first qubit with unitary operation $X^a Z^b$, the second qubit with $X^c Z^d$ and the third qubit with $X^e Z^f$ respectively, here $a,b,c,d,e,f \in \{0,1\}$. Obviously, this encryption process is the same as [Tan and Zhou (2017)]. Our main contribution is to simplify the decryption process (i.e., decryption circuit). To be specific, we get rid of two *SWAP* gates and re-layout the *CNOT* operations, which can be shown between Fig. 3 and Fig. 9. Different from the decryption circuit for *H*, *P*, *CNOT*, the decryption process for *Toffoli* gate is a little complicated, and the extra correction operations *CZ* and *CNOT* are needed. As shown in Fig. 8, considering the first special situation that *f=1*, then the client needs to apply a *CZ* correction between the first and second qubit $TZ_3|\varphi\rangle = CZ_{12}Z_3T|\varphi\rangle$. The second special situation that *c=1*, then the client needs to apply a *CNOT* correction between the first and third qubit $TX_2|\varphi\rangle = CNOT_{13}X_2T|\varphi\rangle$ The third special situation that *a=1*, then the the client needs to apply a *CNOT* correction between the first and third qubits $TX_1|\varphi\rangle = CNOT_{23}X_1T|\varphi\rangle$.

## 4. Correctness and security analysis

### *4.1. Correctness analysis*

In this section, the correctness of the proposed protocol for *Toffoli* gate is verified. Since the correctness of the processes for H, P and CNOT is already verified in [Broadbent (2015)] and [Tan and Zhou (2017)]. Then the only remaining gate is *Toffoli* gate. Assume that the encryption secret keys for the encryption progress of *Toffoli* gate are $(a,b),(c,d),(e,f)$. The verification procedure is given as follows. First, assuming that $b=c=d=e=f=0$ then we can get,

$$T\left(X_1^a \otimes I \otimes I\right)|\varphi\rangle = X^a \otimes CNOT_{23}^a T|\varphi\rangle \tag{3}$$

Assuming that $a=c=d=e=f=0$ then we can get,

$$T\left(Z_1^b \otimes I \otimes I\right)|\varphi\rangle = Z_1^b T|\varphi\rangle \tag{4}$$

Assuming that $a=b=d=e=f=0$ then we can get,

$$T\left(I \otimes X_2^c \otimes I\right)|\varphi\rangle = X_2^c \otimes CNOT_{13}^c T|\varphi\rangle \tag{5}$$

Assuming that $a=b=c=e=f=0$ then we can get,

$$T\left(I \otimes Z_2^d \otimes I\right)|\varphi\rangle = Z_2^d T|\varphi\rangle \tag{6}$$

Assuming that $a=b=c=d=f=0$ then we can get,

$$T\left(I \otimes I \otimes X_2^e\right)|\varphi\rangle = X_2^e T|\varphi\rangle \tag{7}$$

Assuming that $a=b=c=d=e=0$ then we can get,

$$T\left(I \otimes I \otimes Z_3^f\right)|\varphi\rangle = CZ_{12}^f \otimes Z_3^f T|\varphi\rangle \tag{8}$$

Finally, according to equations (3)-(8), we can obtain,

$$T\left(X^a Z^b \otimes X^c Z^d \otimes X^e Z^f\right)|\varphi\rangle = \\ \left(CNOT_{13}^c \otimes X^a Z^b\right)\left(CNOT_{23}^a \otimes X^c Z^d\right)\left(CZ_{12}^f \otimes X^e Z^f\right)T|\varphi\rangle \quad (9)$$

The correctness of the *CZ* correction can be easily verified,

$$CZ\left(X^a Z^b \otimes X^c Z^d\right)|\varphi\rangle = \left(X^a Z^{b+c} \otimes X^{a+c} Z^{b+d}\right)CZ|\varphi\rangle \quad (10)$$

Given that the correctness of encryption and decryption process of *H*, *P*, *T*, *CNOT* and *CZ* has been shown, correctness of the proposed FDQC protocol is obvious: after each round delegation, the client adjusts his secret keys according to Figs. 5-8. So that he can perform the decryption correctly. Because each process of itself is correct, so the proposed FDQC protocol implements the quantum computation as desired.

*4.2 Security analysis*

In the client-server scenario, the security of the proposed protocol contains many aspects, but the main problem is the security of the data and the computation, and as well as the blindness of the message qubits and the delegated quantum operations. The blindness of the message qubits and the delegated quantum operations are discussed in the following parts.

*4.2.1 The blindness of data*

Considering the encryption and decryption processes of *H*, *P*, *CNOT* and *CZ* are the same as Broadbent's DQC protocol, then the processes of the four gates provides the same level of security as the original one, which is perfectly (information-theoretic) secure. Therefore, we will only focus on the security of encrypted qubits which is performing on the encryption and decryption circuit of *Toffoli* gate. Because the client is not able to perform the *CNOT* and *CZ* corrections, then the two operations should be delegated to the server. However, once the server obtains the information of corrections, then the encryption keys of encrypted qubits are exposed (also mentioned in the review of HDQC).

In order to eliminate the particularity of the corrections, the CZ and *CNOT* corrections are added into the fixed order of gates $(H, P, CZ, CNOT, T)$ which is performed in each round delegation. In each round, the server is asked to perform the five unitary operations indistinguishably, therefore there is no mechanism for the server to distinguish the correction operation from the other four operations, so the particularity of the corrections disappears, and the blindness of encryption keys holds. To be specific, suppose that the desired operation is the *Toffoli* gate in one round delegation, and the encryption key *f=1*, then the client will delegate the server to perform the *CZ* correction in the next round delegation. However the *CZ* correction is confused by the other four operations, the server is not able to know that the desired operation is the *CZ* correction, then the security of encryption key f is guaranteed. Since the encrypted qubit which is performing on the gate set $\{H, P, CZ, CNOT, T\}$ is secure, then the blindness of data is obvious.

*4.2.1 The blindness of data*

The computation that the client want to implement can be seen as a desirable circuit which is made up of the delegated quantum operations, therefore the blindness of computation is equivalent to the blindness of the delegated quantum operations. In order to make the delegated quantum operations blind, each operation of the delegated quantum operations is replace by the fixed order of gates $(H, P, CZ, CNOT, T)$, where the *H*, *P*, *CNOT* and *T* operations are needed for the universality, and the *CZ* and *CNOT* operations are needed in the decryption process of certain operation (such as the *Toffoli* gate). Client uses ancillary part to confuse the message part, there is no mechanism for the server to distinguish the message part and the ancillary part, so the server is not able to deduce the desired operations, thus computation that the client wants to implement is blind.

Without loss of generality, we take the delegation of quantum computation *U=HP* as an example. If the client wants to ask the server to perform the *U* on the encrypted qubit $X^a Z^b |\varphi\rangle$, then the whole produce of FDQC is as follows, he firstly generates a 9-qubit sequence $S1$ which consists of ancillary qubits and message qubits, where the message qubit is in the subsequence $S_p$ (i.e., the message part), and the other four subsequence $S_H$, $S_{CZ}$, $S_{CNOT}$ and $S_T$ are the ancillary part. The client sends $S1$ to the sever to perform the fixed order of gates. Because the server cannot distinguish the message part from the ancillary part, he cannot know that the desired operation which is performed on the message qubit is the *P* gate, so in this round, the delegated quantum operation *P* is secure. Then the server sends all the qubits back to the client, the client reconstructs the message part and the ancillary part(the ancillary qubits can be reused) according to the delegated operation *H* in next round. He sends new generated $S2$ to the server to implement the fixed order of gates again, the server still cannot know that the desired operation is the *H* gate in this round. Finally, he sends all the qubits back to client, and when the computation $HPX^a Z^b |\varphi\rangle$ is done, the client decrypts $HPX^a Z^b |\varphi\rangle$ according to the decryption rules. During the process, the server cannot learn anything about client's desired operations, thus, the computation is blind.

## 5. Conclusion

As quantum devices are scarce and expensive, it is not hard to imagine that very few companies or scientific institutions can have a quantum device or a quantum computer in the foreseeable future. It is an impossible mission for the quantum computer to be popularized in the following decades. But the delegating private quantum computation provide a solution, which will enables the ordinary client with uncomplicated quantum device to perform the quantum computation with unconditional security. Thus delegating quantum computation to the remote server has strong practical and economic motivation. In recent years, quite a lot delegating private computation protocols have been proposed, but some protocols might exist the design flaws that might cause some security problems. Therefore the improvement of the existing protocols is also an attractive work.

In this study, we pointed out that the decryption circuit of *Toffoli* gate is a little complicated and the information leaking risk exists in HDQC. For solving the problem of protecting the blindness of computation, we propose an full-blind DQC protocol with $\{H, P, CNOT, T\}$. In the practical application, the quantum gate set $\{H, P, CNOT, T\}$ seems more commonly used than $\{X, Z, H, P, R, CNOT\}$ as the *Toffoli* gate (i.e., *T*) is considered to be the basic unit for constructing complex quantum circuits. So the research on DQC with $\{H, P, CNOT, T\}$ is a meaningful work. In the proposed protocol, although we have optimized the decryption circuit of

*Toffoli* gate, it still needs multiple interactions. One of our future work is to further simplify the decryption circuit of *Toffoli* gate, and reduce the times of interaction, even get rid of the interaction.

As FDQC can provide a secure "client-server" mode for universal quantum computation, one hand, we can try to use this model to solve some classic security calculation problems [Pradeep (2016); Cao (2018); Liu (2018)]. Another another important work is to combine DQC with some practical quantum protocols, such as quantum key agreement [Liu (2017); Liu (2018)], quantum private comparison [Yang (2009); Liu (2014a); Liu (2014b); Liu (2014c)], quantum sealed-bid Auction [Liu (2014d); Liu (2016)], which will be another interesting direction to be further studied. We conclude this paper with an expectation that the works reported here will be realized experimentally and further applied in the daily life.

**Acknowledgement**: The authors would like to thank the anonymous reviewers and editor for their comments that improved the quality of this paper. This work is supported by the National Nature Science Foundation of China (Grant Nos. 61502101 and 61501247), the Natural Science Foundation of Jiangsu Province, China (Grant No. BK20171458), the Six Talent Peaks Project of Jiangsu Province, China (Grant No. 2015-XXRJ-013), the Natural science Foundation for colleges and universities of Jiangsu Province, China (Grant No. 16KJB520030), the Research Innovation Program for College Graduates of Jiangsu Province, China (Grant No. KYCX17_0902), the Practice Innovation Training Program Projects for the Jiangsu College Students (Grant No. 201810300016Z), and the Priority Academic Program Development of Jiangsu Higher Education Institutions (PAPD).

**References**
**Aharonov, D.; Ben-Or, M.; Eban, E.** (2008): Interactive proofs for quantum computations. In *Proceedings of Innovations in Computer Science*, pp. 453-469.
**Ambainis, A.; Mosca, M.; Tapp, A.; De Wolf, R.** (2000): Private quantum channels. In *41rd Symposium on Foundations of Computer Science*, pp. 547-553.
**Arrighi, P.; Salvail, L.** (2006): Blind quantum computation. *International Journal of Quantum Information*, vol. 4, no. 5, pp. 883-898.
**Boykin, P. O.; Roychowdhury, V.** (2003): Optimal encryption of quantum bits. *Physical Review A*, vol. 67, no. 4, pp. 645-648.
**Broadbent, A.; Fitzsimons, J.; Kashefi, E.** (2008): Universal blind quantum computation. *In 2009 50th Annual Ieee Symposium on Foundations of Computer Science*, pp. 517-526.
**Broadbent, A.; Gutoski, G.; Stebila, D.** (2013): Quantum one-time programs. In *Advances in Cryptology – CRYPTO 2013*. vol. 8043, no. Suppl 1, pp. 344-360.
**Broadbent, A.** (2015): Delegating private quantum computations. *Canadian Journal of Physics*, vol. 93, no. 9, pp. 410-413.
**Cao, Y.; Zhou, Z. L.; Sun, X. M.; Gao, C. Z.** (2018): Coverless information hiding based on the molecular structure images of material. *CMC-Computers Materials & Continua*, vol. 54, no. 2, pp. 197-207.
**Childs, A. M.** (2005): Secure assisted quantum computation. *Rinton Press, Incorporated*.
**Fisher, K. A. G.; Broadbent, A.; Shalm, L. K.; Yan, Z.; Lavoie, J.; Prevedel, R.;Jennewein, T.; Resch, K. J.** (2013): Quantum computing on encrypted data. *Nature Communications*, vol. 5, no. 2, pp. 3074.

12**Gentry; Craig** (2009): Fully homomorphic encryption using ideal lattices. *In 41rd annual ACM symposium on Theory of computing*, vol. 9, no. 4, pp. 169-178.

**Kong, X.; Li, Q.; Wu, C.; Yu, F.; He, J.; Sun, Z.** (2016): Multiple-server flexible blind quantum computation in networks. *International Journal of Theoretical Physics*, vol. 55, no. 6, pp. 3001-3007.

**Li, Q.; Chan, W. H.; Wu, C.; Wen, Z.** (2014): Triple-server blind quantum computation using entanglement swapping. *Physical Review A*, vol. 89, no. 4, pp. 2748-2753.

**Liu, W.-J.; Chen, Z.-Y.; Ji, S.; Wang, H.-B.; Zhang, J.** (2017): Multi-party semiquantum key agreement with delegating quantum computation. *International Journal of Theoretical Physics*, vol. 56, no. 10, pp. 3164-3174.

**Liu, W.-J.; Xu, Y.; Yang, C.-N.; Gao, P.-P.; Yu, W.-B.** (2018): An efficient and secure arbitrary n-party quantum key agreement protocol using bell states. *International Journal of Theoretical Physics*, vol. 57, no. 1, pp. 195-207.

**Liu, W. J.; Liu, C.; Liu, Z. H.; Liu, J. F.; Geng, H. T.** (2014a): Same initial states attack in yang et al.'s quantum private comparison protocol and the improvement. *International Journal of Theoretical Physics*, vol. 53, no. 1, pp. 271-276.

**Liu, W. J.; Liu, C.; Chen, H. W.; Li, Z. Q.; Liu, Z. H.** (2014b): Cryptanalysis and improvement of quantum private comparison protocol based on bell entangled states. *Communications in Theoretical Physics*, vol. 62, no. 8, pp. 210-214.

**Liu, W. J.; Liu, C.; Wang, H. B.; Liu, J. F.; Wang, F.; Yuan, X. M.** (2014c): Secure quantum private comparison of equality based on asymmetric w state. *International Journal of Theoretical Physics*, vol. 53, no. 6, pp. 1804-1813.

**Liu, W. J.; Wang, F.; Ji, S.; Qu, Z. G.; Wang, X. J.** (2014d): Attacks and improvement of quantum sealed-bid auction with epr pairs. *Communications in Theoretical Physics*, vol. 61, no. 6, pp. 686.

**Liu, W. J.; Wang, H. B.; Yuan, G. L.; Xu, Y.; Chen, Z. Y.; An, X. X.; Ji, F. G.; Gnitou, G. T.** (2016): Multiparty quantum sealed-bid auction using single photons as message carrier. *Quantum Information Processing*, vol. 15, no. 2, pp. 869-879.

**Liu, Y.; Peng, H.; Wang, J.** (2018): Verifiable diversity ranking search over encrypted outsourced data. *Cmc-Computers Materials & Continua*, vol. 55, no. 1, pp. 037-057.

**Morimae, T.** (2012a): Continuous-variable blind quantum computation. *Physical Review Letters*, vol. 109, no. 23, pp. 230502.

**Morimae, T.; Fujii, K.** (2012b): Blind topological measurement-based quantum computation. *Nature Communications*, vol. 3, no. 3, pp. 1036.

**Morimae, T.; Dunjko, V.; Kashefi, E.** (2015): *Ground state blind quantum computation on AKLT state*. Rinton Press, Incorporated.

**Nielsen, M. A.; Chuang, I. L.** (2011): *Quantum Computation and Quantum Information: 10th Anniversary Edition*. Cambridge University Press.

**Pradeep, A.; Mridula, S.; Mohanan, P.** (2016): High security identity tags using spiral resonators. *CMC-Computers Materials & Continua*, vol. 52, no. 3, pp. 187-196.

**Rivest, R. L.; Adleman, L.; Dertouzos, M. L.** (1978): On data banks and privacy homomorphisms. *Foundations of Secure Computation*, pp. 169-179.

**Tan, X.; Zhou, X.** (2017): Universal half-blind quantum computation. *Annals of Telecommunications*, vol. 72, no. 9-10, pp. 1-7.

**Xu, H. R.; Wang, B. H.** (2014): Universal single-server blind quantum computation for classical client. *Eprint Arxiv*:1410.7054.

**Yang, Y. G.; Wen, Q. Y.** (2009): An efficient two-party quantum private comparison protocol with decoy photons and two-photon entanglement. *Journal of Physics A-Mathematical and Theoretical*, vol. 42, no. 5. pp. 055305